\documentclass[amssymb,prl,
twocolumn,
,citeautoscript,floatfix,
footinbib,superscriptaddress]{revtex4}
\usepackage{graphicx}
\usepackage[latin1]{inputenc}
\usepackage{times}
\usepackage{mathptm}
\newcommand\beq{\begin{equation}}
\newcommand\eeq{\end{equation}}
\newcommand\bea{\begin{eqnarray}}
\newcommand\eea{\end{eqnarray}}

\newcommand\ben{\begin{enumerate}}
\newcommand\een{\end{enumerate}}

\newcommand\TTS{Sr$_3$Ru$_2$O$_7$}

\begin{document}

\title{Quantum oscillations in the anomalous phase in \TTS}

\author{J.-F. Mercure}
\affiliation{Scottish Universities Physics Alliance (SUPA), School of Physics and Astronomy, University of St Andrews, North Haugh, St Andrews KY16 9SS, United Kingdom}

\author{S. K. Goh}
\affiliation{Cavendish Laboratory, University of Cambridge, J.J. Thomson Avenue, Cambridge CB3 0HE, United Kingdom}

\author{E. C. T. O'Farrell}
\affiliation{Cavendish Laboratory, University of Cambridge, J.J. Thomson Avenue, Cambridge CB3 0HE, United Kingdom}

\author{R. S. Perry}
\affiliation{SUPA, School of Physics, University of Edinburgh, Mayfield Road, Edinburgh EH9 3JZ, United Kingdom}

\author{M. L. Sutherland}
\affiliation{Cavendish Laboratory, University of Cambridge, J.J. Thomson Avenue, Cambridge CB3 0HE, United Kingdom}

\author{A. W. Rost}
\affiliation{Scottish Universities Physics Alliance (SUPA), School of Physics and Astronomy, University of St Andrews, North Haugh, St Andrews KY16 9SS, United Kingdom}

\author{S. A. Grigera}
\affiliation{Scottish Universities Physics Alliance (SUPA), School of Physics and Astronomy, University of St Andrews, North Haugh, St Andrews KY16 9SS, United Kingdom}
\affiliation{Instituto de F\'isica de liquidos y sistemas biologicos, UNLP, 1900 La Plata, Argentina.}

\author{R. A. Borzi}
\affiliation{Instituto de Investigaciones Fisicoqu\'imicas Te\'oricas y Aplicadas (UNLP-CONICET), 
c.c. 16, Suc. 4 and Departamento de F\'isica, IFLP, UNLP, c.c. 
67, 1900 La Plata, Argentina.}
 
\author{P. Gegenwart}
\affiliation{I. Physik. Institut, Georg-August-Universit$\ddot{a}$t G$\ddot{o}$ttingen, D-37077 G$\ddot{o}$ttingen, Germany}

\author{A. P. Mackenzie}
\affiliation{Scottish Universities Physics Alliance (SUPA), School of Physics and Astronomy, University of St Andrews, North Haugh, St Andrews KY16 9SS, United Kingdom}

\begin{abstract}
We report measurements of quantum oscillations detected in the putative nematic phase of \TTS. Significant improvements in sample purity enabled the resolution of small amplitude dHvA oscillations between two first order metamagnetic transitions delimiting the phase. Two distinct frequencies were observed, and their amplitudes follow the normal Lifshitz-Kosevich profile. The Fermi surface sheets seem to correspond to a subset of those detected outside the phase. Variations of the dHvA frequencies are explained in terms of a chemical potential shift produced by reaching a peak in the density of states, and an anomalous field dependence of the oscillatory amplitude provides information on domains.
\end{abstract}

\maketitle


In recent years, the possibility that electronic liquid crystalline states might exist~\cite{Kivelson} has generated considerable interest. In such systems, the electron fluid is thought to develop textures leading to anisotropies which are not compatible with the space group symmetry of the host crystal. Experimentally, nematic-like transport properties have been observed in high-purity two dimensional electron gases~\cite{Lilly, Lilly2, Pan, Cooper}, and signatures of electronic anisotropy have also been reported in YBa$_2$Cu$_3$O$_{7-\delta}$~\cite{Ando, Hinkov}.   Combined with observations on the subject of this letter, \TTS, these experiments have stimulated a body of further theoretical work on the issue~\cite{Dell'Anna, Oganesyan, Fradkin, yamase, Kee, Kee2, PRLGreen, Quintanilla, Berridge, ISI:000264380200049, ISI:000267699200063, lee-2009}.  

\TTS\ offers an excellent opportunity for experimental study of this novel behaviour. In the cleanest single crystals, with mean free paths of several thousand angstroms, a well-defined phase displaying nematic-like transport anisotropies exists below 1.2~K for magnetic fields between approximately 7.9 and 8.1~T~\cite{perry2, science2, science3}.  The combination of long mean free paths and the existence of large single crystals means that it is possible to obtain thermodynamic as well as transport information about this phase and its formation~\cite{science4}.   

In this letter, we present a study of the nematic phase in \TTS\ using one of the best-established microscopic probes of metals, the de Haas - van Alphen (dHvA) effect.  Performing the project required the growth of a new generation of high purity single crystals, screened in a series of dHvA measurements to provide the largest oscillatory signals.  This enabled the observation of oscillations both within the nematic phase itself and in its immediate vicinity.

The crystals were grown using methods published previously~\cite{perryGrowth}. Measurements with a noise floor of around 30~pV/$\sqrt{\textrm{Hz}}$ were performed in Cambridge using a dilution refrigerator. An improvement in signal due to crystal purity of a factor of approximately 25 was obtained compared to the previous dHvA study~\cite{borzi,Note2,MercurePRBPreprint}. 

\begin{figure}[t]
  \centering
	\includegraphics[width=1\columnwidth]{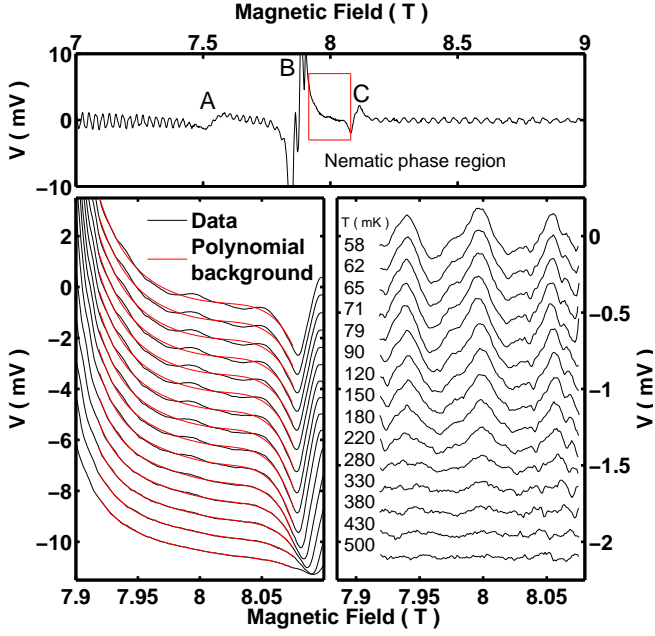}
	\caption{dHvA oscillations in the nematic phase. $Top$ dHvA oscillations in $\partial^2 M / \partial H^2$ in the region surrounding the metamagnetic transition. From left to right, three magnetic features are visible, denoted $A$, $B$ and $C$ on the diagram (see text). $Left$ The signal in the nematic phase region, along with fifth order polynomial background subtraction, shown in red. $Right$ Nematic phase oscillations remaining after background removal at 15 different temperatures. In the left and right graphs, curves are offset for clarity, and include a total amplification factor of 10$^5$ from low temperature transformers and preamplifiers.}
	\label{fig:Oscillations}
\end{figure}

In Fig. \ref{fig:Oscillations} we present plots of the dHvA oscillations, inside and outside the nematic phase, with the field oriented along the $c$-axis. Second harmonic detection was used, in order to reduce the large non-oscillatory component of the susceptibility near the metamagnetic transition. Consequently, metamagnetic jumps in the magnetisation appear as sharp asymmetric peaks. The top panel of Fig. \ref{fig:Oscillations} shows the dHvA signal in the region neighbouring the metamagnetic transition, where three such peaks can be observed, denoted $A$, $B$ and $C$. The broad peak $A$ located at 7.5~T corresponds to a metamagnetic cross-over previously observed in several properties \cite{perry2}, while the sharp features $B$ and $C$ correspond to metamagnetic phase transitions at 7.9 and 8.1~T, bounding the nematic phase. 

In the bottom left panel of Fig.~\ref{fig:Oscillations}, we present a blow-up of the second harmonic susceptibility signal between 7.9 and 8.1~T at 15 different temperatures between 58 and 500~mK, offset for clarity. Small amplitude oscillations are observed on top of a smooth magnetic background, which was modelled by fitting a fifth order polynomial (red line) to the data between 7.915 and 8.080~T. Polynomials of lower order did not reproduce the magnetic background well, while higher orders removed a component of the oscillations. In Fig.~\ref{fig:Oscillations}, right, we show the oscillations remaining after the background removal. These are suppressed as the temperature increases to 500~mK, due to the Lifshitz-Kosevich (LK) amplitude reduction factor~\cite{shoenberg}. The quality of the LK fits suggests that both orbits result from Landau quasiparticles. The oscillations were reproduced on three samples over four different dHvA runs. They were moreover found to be suppressed by rotating the field as little as 5$^{\circ}$ from $c$-axis.

\begin{figure}[t]
  \centering
	\includegraphics[width=0.945\columnwidth]{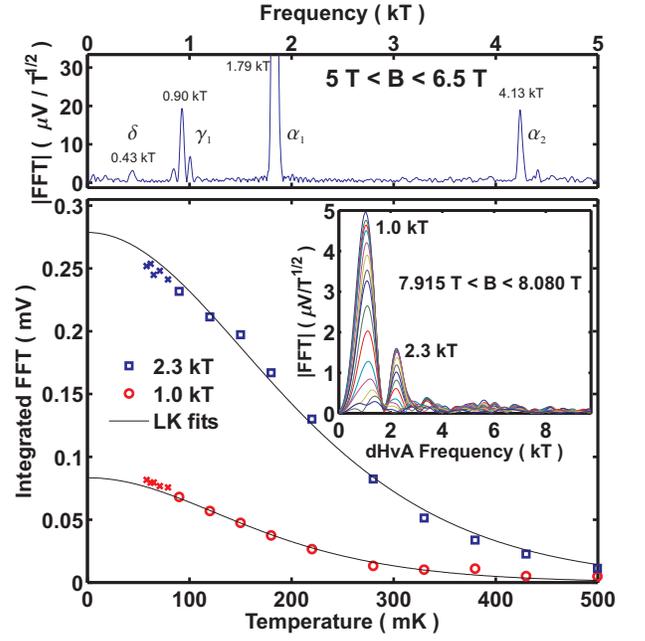}
	\caption{$Top$ Fourier spectrum of the dHvA data on the low field side of the transition, between 5 and 6.5~T, with the field aligned with the $c$-axis. $Inset$ Fourier spectra of the nematic phase data at all temperatures. $Main$ $panel$ Temperature dependence of the amplitude of the nematic phase oscillations. Solid lines represent LK fits to the data, which revealed quasiparticle masses of $6.1 \pm 0.3$~$m_e$ and $7.4 \pm 0.2$~$m_e$ for the 1.0 and 2.3~kT peaks respectively. Data plotted with crosses correspond to temperatures below 90~mK which were corrected, as explained in \cite{Note1}, and were not used in the non-linear fits.}
	\label{fig:LKInset1}
\end{figure}

\begin{figure}[t]
  \centering
	\includegraphics[width=0.885\columnwidth]{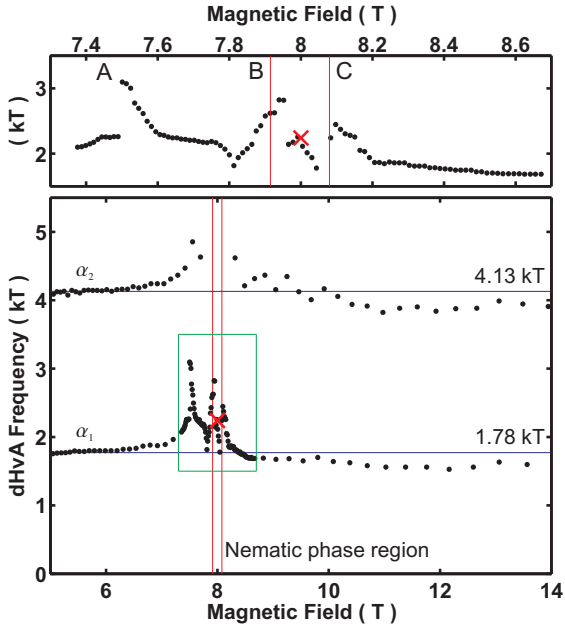}
	\caption{Field dependence of the dHvA frequencies situated near 1.8 and 4.1~kT in the low field side. These feature non-linear changes in the metamagnetic region. The data points correspond to the location of peaks in FFTs taken over field windows of equal inverse field width of 0.005~T$^{-1}$, and the field value for each data point corresponds to the inverse of the average inverse field. The red lines represent the nematic phase field boundaries, and the red cross the value of frequency measured inside the nematic phase. The top panel is a blow-up of the data within the green square in the bottom panel. Three peaks in $F(B)$, denoted $A$-$C$ are observed near each of the metamagnetic features shown in Fig.~\ref{fig:Oscillations}. Corrections to field values due to the intrinsic magnetisation were found to be small.}
	\label{fig: FreqCurves6}
\end{figure}

The oscillatory data taken in the nematic phase revealed two distinct frequencies of 1.0 and 2.3~kT (inset to Fig.~\ref{fig:LKInset1}), with masses of 6.1$\pm$0.3~$m_e$ and 7.4$\pm$0.2~$m_e$ respectively~\cite{Note1}. For reference, we show, in the top panel of Fig.~\ref{fig:LKInset1}, a spectrum obtained outside the nematic phase, between 5 and 6.5~T. Peaks are observed at 0.43, 0.90, 1.78 and 4.13~kT termed $\delta$, $\gamma_1$, $\alpha_1$ and $\alpha_2$ respectively~\cite{tamai}; higher resolution studies of the low frequency region reveal further frequencies, termed $\beta$ and $\gamma_2$, at 0.15~kT and 0.11~kT. The associated quasiparticle masses rule out the possibility that any of these frequencies is a harmonic of any other (see \cite{tamai, borzi, MercurePRBPreprint} for details). The resolution with which we can determine the frequency of the 1.0~kT peak in the nematic region is limited by the number of oscillations present in the narrow field range, and it seems reasonable to associate it with the 0.9~kT peak seen on the low field side. At first sight, it is difficult to relate the 2.3~kT peak with those seen at low fields, but, as shown in Fig.~\ref{fig: FreqCurves6}, the $\alpha_1$ and $\alpha_2$ frequencies show a strong field dependence in the metamagnetic region, and the 2.3~kT frequency is compatible with an interpolation into the nematic phase of the field dependence of the $\alpha_1$ frequency. This identification is also compatible with mass analysis (presented in detail in ref.~[\onlinecite{MercurePRBPreprint}]) which shows that the masses associated with the frequencies in the nematic phase are similar to those for the $\beta$ and $\alpha_1$ frequencies (6.6$\pm$1~$m_e$ and 7.0$\pm$0.5~$m_e$ respectively) outside it.

\begin{figure}[t]
  \centering
	\includegraphics[width=1\columnwidth]{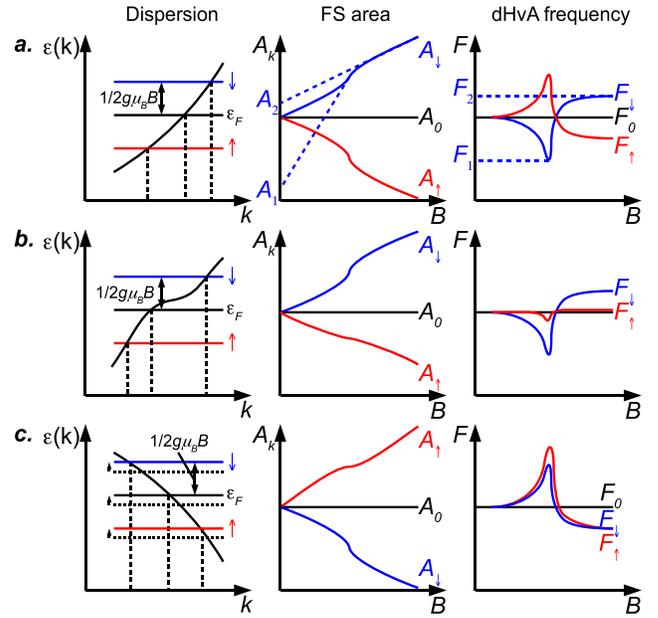}
	\caption{Three scenaria where field dependent dHvA frequencies may be encountered. In each case, the dispersion is shown on the left, the Fermi surface cross-sectional area as a function of magnetic field, $A(B)$, in the centre and the dHvA frequency as a function of magnetic field, $F(B)$, on the right. Large changes in the slope of $A(B)$ lead to peaks in $F(B)$~\cite{JulianUPt3}. $a)$ A field dependent $g$ factor leads to non-linear changes in $A(B)$ for each spin species, producing peaks of opposite sign in $F(B)$. $b)$ A flattening of the dispersion leads to a large change in $F(B)$ for only one spin species. $c)$ A change in chemical potential leads to an acceleration of the change in $A(B)$ for one spin species and a deceleration for the other. This leads to peaking of $F(B)$ that is similar for both spin species.}
	\label{fig: BackProjectionPaper4}
\end{figure}

The dependence of the frequency $F$ of dHvA oscillations in the vicinity of metamagnetism encodes information about the mechanism of the transition. As pointed out by Julian and co-workers~\cite{JulianUPt3}, the measured dHvA frequency $F(B)\propto A(B)-B\mathrm{d}A(B)/\mathrm{d}B$, where $A(B)$ is the Fermi surface cross-sectional area as a function of magnetic field $B$. The situation most usually considered, sketched in Fig.~\ref{fig: BackProjectionPaper4}$a$, is that where the coupling between field and magnetisation is field dependent (e.g. the effective $g$ factor is field dependent), and leads to symmetrical changes in the cross-sectional area of the Fermi surface for both spins around the zero field value. This produces peaks in the measured frequency of opposite sign.

The situation in Fig.~\ref{fig: BackProjectionPaper4}$b$ originates from a pronounced change of gradient in the dispersion of the band being studied, which may appear either below or above the Fermi level. Such a dispersion leads to changes in mainly one of the spin split cross-sectional areas, and produces a peak in one of the measured frequencies, leaving the other almost constant. Large frequency splitting is therefore expected through the transition region and at fields above it. 

Finally, the third scenario depicted in Fig.~\ref{fig: BackProjectionPaper4}$c$ is that in which a change in chemical potential occurs at a certain field value. This would correspond to a sudden dispersion change in one of the other bands in a multi-band material like \TTS, leading to a redistribution of carriers. This chemical potential shift leads to an acceleration of the splitting of the cross-sectional area for one spin, and to a slowdown for the other. Consequently, peaks of the same sign for both spins are observed in the measured frequencies, and both saturate at high-field values reduced from those on the low-field side~\cite{Note7}. 

Only the third scenario is compatible with our observations. We see peaking of both the $\alpha_1$ and $\alpha_2$ frequencies through the metamagnetic region, followed by saturation at constant values of approximately 1.6 and 3.9~kT respectively, but do not resolve any frequency splitting through the transition (see Figs. \ref{fig: FreqCurves6} and \ref{fig: BackProjectionPaper4}c). 

This finding is consistent with the results of Tamai $et$ $al.$ \cite{tamai}, who showed that a saddle point in the dispersion of $\gamma_2$ is present below the Fermi level ($E_\mathrm{F}$). This feature of the band structure leads to a peak in the density of states which is situated at around 3 meV below $E_\mathrm{F}$. Conservation of the number of electrons requires that when approaching this saddle point with Zeeman splitting, the hole filling of the peak in the density of states by one of the spin species leads to an increase of the chemical potential compared to its zero field value. This corresponds to an exchange of holes from existing bands towards the region of the Brillouin zone with a large and expanding density of states, and produces the observed increase in magnetisation. Both $\alpha_1$ and $\alpha_2$ correspond to hole sheets of the Fermi surface, so this is consistent both with the reduction of their dHvA frequencies at high fields (above 8.1~T) and with the positive sign of the peaks in the frequencies through the metamagnetic region.

This observation has interesting implications for understanding both the metamagnetism and the effects that metamagnetic fluctuations have on different bands near $E_F$. The $\alpha_1$ and $\alpha_2$ sheets are based on the $d_{xz}$ and $d_{yz}$ orbitals of Ru and hence have a more 1-D character than the other Fermi surface sheets. It appears that these bands contribute only weakly to the magnetic moment change, but still couple to the metamagnetism via the charge transfer that changes the chemical potential. This in turn implies a sensitivity of the $\alpha_1$ and $\alpha_2$ sheets to the fluctuations that occur as one or more other bands (most likely $\gamma_2$) approach metamagnetic criticality.

Having observed well defined orbits in the nematic region we can analyze changes in the amplitude of each frequency to investigate microscopic scattering on different Fermi surface sheets. We first consider the normal region; in the top panel of Fig.~\ref{fig:LKInset1}, the relative amplitude of the $\beta$ to the $\alpha_1$ and $\alpha_2$ signals is 1~:~12~:~1.3. The mean free path determined from a Dingle analysis that describes damping of quantum oscillations by a homogeneous distribution of impurities is 270~nm, fully consistent with that deduced from the resistivity using the Fermi surface from ref. \cite{tamai}. Going into the nematic region, a naïve analysis of the resistivity would suggest that the mean free path halves~\cite{perry2}. Using the Dingle factor and taking into account the Bessel function factor associated with the field modulation technique~\cite{Note3} we can calculate relative amplitudes for a reduced mean free path of 135~nm by extrapolating the amplitudes from just outside this region. This gives an amplitude ratio 1~:~8~:~0.5. The observed ratio (inset Fig.~\ref{fig:LKInset1} main panel) is very different: 1~:~0.4~:~0. This strongly indicates that the assumption of uniform impurity scattering is no longer valid in the nematic region.

To explain this discrepancy we consider the existence of domains as discussed in refs. \cite{science2} and \cite{science3}. In those studies we commented that the rise in resistivity in the nematic region for fields applied parallel to the $c$-axis and the transport anisotropy seen in slightly tilted fields were both consistent with strong scattering at domain walls. For quantum oscillations this hypothesis implies that only frequencies due to orbits entirely within a single domain survive; therefore, the observed orbits can be used as real space callipers of the domain's smallest linear dimension. The cyclotron diameters of the $\beta$, $\alpha_1$ and $\alpha_2$ orbits are 270~nm, 385~nm and 580~nm respectively~\cite{Note4}, so the strong frequency dependence of the signal suppression suggests that the sample contains a distribution of domains with an average wall separation of order 500~nm. Estimating their linear dimension is an important step forward, and should motivate studies with probes capable of imaging them directly in real space.

The above analysis gives clues about the microscopic nature and mechanism of formation of the novel phase in \TTS, but the most important finding presented here is the discovery that quantum oscillations can be observed within it. Combined with the observation that the frequencies change across the metamagnetic region (Fig.~\ref{fig: FreqCurves6}) and recent Hall effect measurements showing that the Hall number does not drop on entry to the phase~\cite{Borzi09}, our work provides strong evidence that the phase diagram of \TTS\ is characterised by a series of fluid-fluid transitions. Further, the fluid in the nematic phase contains Landau Fermi liquid quasiparticles (Fig.~\ref{fig:LKInset1}) whose quantum oscillatory amplitude conforms to the prediction of the Lifshitz-Kosevich theory. Since dHvA is sensitive to Fermi surface areas and not to shape, this set of observations is compatible with any microscopic model postulating that the root of the nematic transport properties lies in a Pomeranchuk-like Fermi surface distortion.

In conclusion, we have succeeded in detecting quantum oscillations in the nematic phase of \TTS. Our results establish that at least some of the mobile charges in the phase are Landau quasiparticles, and also reveal the importance of a chemical potential shift and domain physics in understanding the phase's formation and properties. 

We thank A. M. Berridge, C. A. Hooley and G. G. Lonzarich for informative discussions, and A. S. Gibbs for providing crystals used to determine the conditions for sample equilibrium~\cite{Note1}. This work was supported by the Engineering and Physical Sciences Research Council. 

\bibliographystyle{aps3etal}

\end{document}